\documentclass[12pt,letterpaper]{article}
\pdfoutput=1
\usepackage[top=3cm, bottom=3cm, left=2cm, right=2cm]{geometry}
\usepackage[usenames,dvipsnames,svgnames]{xcolor}
\definecolor{darkblue}{rgb}{0.0,0.1,0.3} % dark blue
\definecolor{darkgreen}{rgb}{0,0.65,0}
\definecolor{dblue4}{rgb}{0.06,0.31,0.55} % DodgerBlue4
\definecolor{nicered}{rgb}{0.7,0.1,0.1}
\definecolor{nicegreen}{rgb}{0.1,0.5,0.1}

\usepackage[numbers,sort&compress]{natbib}
\usepackage[utf8x]{inputenc}
\usepackage{amsmath,amssymb,amsfonts,amsthm}
\usepackage{xcolor}
\usepackage[colorlinks=true,citecolor= nicegreen,linkcolor=nicered]{hyperref}

\usepackage{verbatim}
\usepackage{graphicx}
\graphicspath{{figures/}}
\usepackage{cancel}
\usepackage{tipa}
\usepackage{array}   % for \newcolumntype macro
\newcolumntype{L}{>{$}l<{$}} % math-mode version of "l" column type
\newcolumntype{R}{>{$}r<{$}} % math-mode version of "l" column type
\usepackage{booktabs}
\usepackage{tabularx}
\usepackage{multirow}
\usepackage{dsfont}
\newcolumntype{Y}{>{\centering\arraybackslash}X}
\usepackage{tikz}
\newcommand{\ReportNumbers}[1]{%
\begin{tikzpicture}[overlay, remember picture]
\path (current page.north east) ++(-1,-1) node[below left] {#1};
\end{tikzpicture}
}

\usepackage{comment}
\includecomment{details}
\specialcomment{details}
{\begingroup}{\endgroup}

\title{Anomaly-free Abelian gauge symmetries\\with Dirac scotogenic models}

\author{Nicolás Bernal\footnote{\href{mailto:nicolas.bernal@uan.edu.co}{nicolas.bernal@uan.edu.co}}\\
\textit{\small  Centro de Investigaciones, Universidad Antonio Nariño}\\
\textit{\small  Carrera 3 Este \# 47A-15, Bogotá, Colombia}\\
[4mm]
Julián Calle\footnote{\href{mailto:julian.callem@udea.edu.co}{julian.callem@udea.edu.co}}\,
and Diego Restrepo\footnote{\href{mailto:restrepo@udea.edu.co}{restrepo@udea.edu.co}}\\
\textit{\small Instituto de Física, Universidad de Antioquia}\\
\textit{\small  Calle 70 \# 52-21, Apartado Aéreo 1226, Medellín, Colombia}
}
\date{}
\begin{document}
\maketitle
\ReportNumbers{\footnotesize PI/UAN-2021-685FT}
\begin{abstract}
We perform a {\it systematic} analysis of standard model extensions with an additional anomaly-free gauge $U(1)$ symmetry, to generate Dirac neutrino masses at one loop.
Under such symmetry, standard model fields could either transform or be invariant, corresponding to an active $U(1)_X$ or a dark $U(1)_D$ symmetry, respectively.
Having an anomaly-free symmetry imposes nontrivial conditions to the number and charges of the unavoidable new states.
We perform an intensive scan, looking for non-anomalous solutions for given number of extra chiral fermions.
In particular, we concentrate on solutions giving rise to scotogenic neutrino masses via the effective Dirac mass operator.
We study the cases where the Dirac mass operator with dimension 5 or 6, is mediated by Dirac or Majorana states, and corresponds to an active $U(1)_X$ or a dark $U(1)_D$ symmetry.
Finally, we comment on the solutions featuring no massless chiral fermions.
\end{abstract}

%%%%%%%%%%%%%%%%%%%%%%%%%%%%%%%%%%%%%%%%%%%%%%%%%%%%%%%%%%%%
%%%%%%%%%%%%%%%%%%%%%%%%%%%%%%%%%%%%%%%%%%%%%%%%%%%%%%%%%%%%
%%%%%%%%%%%%%%%%%%%%%%%%%%%%%%%%%%%%%%%%%%%%%%%%%%%%%%%%%%%%
\section{Introduction}

The standard model (SM) of particle physics is a very successful theory, even though it has to be extended in order to account for neutrino masses and dark matter (DM).
The interpretation of neutrino experimental data in terms of neutrino oscillations is compatible with both Majorana or Dirac neutrino masses~\cite{Zyla:2020zbs}, with no theoretical preference for either of the possibilities.
However, most of the proposals in the literature assume that neutrinos are Majorana in nature (see Ref.~\cite{Cai:2017jrq} for a review), whereas mass generation mechanisms for Dirac type neutrinos are less studied but have recently received increased attention (see, i.e., Refs.~\cite{Sayre:2005yh, Batra:2005rh, Nakayama:2011dj, Heeck:2012bz, Ma:2014qra, deGouvea:2015pea, Ma:2015mjd, Bonilla:2016zef, Chulia:2016ngi, Chulia:2016giq, Bonilla:2016diq, Ma:2016mwh, Wang:2016lve, Wang:2017mcy, Borah:2017dmk,
Hirsch:2017col,
Yao:2017vtm,
CentellesChulia:2017koy,
Yao:2018ekp, Reig:2018mdk, Han:2018zcn, Bonilla:2018ynb, Calle:2018ovc,
Borah:2018gjk,
Borah:2018nvu,
Carvajal:2018ohk,
CentellesChulia:2018gwr,
Saad:2019bqf, Jana:2019mez, Calle:2019mxn, Jana:2019mgj,
Enomoto:2019mzl,
Borah:2019bdi,
Restrepo:2019soi,
Dasgupta:2019rmf,
Ma:2019iwj,
Ma:2019byo,
CentellesChulia:2020dfh,
Wang:2020dbp, Wong:2020obo, Ma:2021szi}).

To explain Dirac neutrino masses, right-handed neutrinos (RHNs) have to be introduced, however, that is not enough.
An extra local
symmetry is also required to guarantee proper total lepton number
conservation~\cite{Ma:2014qra}.  Even so, the resulting Yukawa couplings can
turn out to be too small, of the order $\mathcal{O}\left(10^{-10}\right)$ or even smaller, if Dirac neutrino masses are induced directly
from the SM Higgs mechanism~\cite{Calle:2018ovc, CentellesChulia:2019gic}.  Nevertheless, if the symmetry forbids the tree-level contribution driven by the SM Higgs, a Dirac-seesaw mechanism can be implemented.
For example, the type-I Dirac-seesaw could appear in the context of anomaly-free gauge $U(1)_{B-L}$ symmetries~\cite{Ma:2014qra}.
At one-loop, the heavy particles
in the radiative seesaw can be fully associated to an Abelian gauge dark symmetry $U(1)_D$ with the lightest of them as DM candidate~\cite{Gu:2007ug, Batell:2010bp, Farzan:2012sa, Calle:2019mxn}, as well to an active Abelian gauge symmetry $U(1)_X$, like an $U(1)_{B-L}$~\cite{Calle:2018ovc, Bonilla:2018ynb, Calle:2019mxn}.  Until now, the studies of one-loop Dirac neutrino masses have typically focused on finding specific anomaly-free solutions of this two kinds of symmetries, see, e.g., Refs.~\cite{Saad:2019bqf, Bonilla:2019hfb, Jana:2019mez, Jana:2019mgj, Escribano:2020iqq, Wong:2020obo}.
Here, we present a complete set of relevant anomaly-free solutions to the
general problem of the generation of Dirac neutrino masses at
one-loop with chiral singlet-fermions.
Each of the solutions leads to a unique model with
its specific phenomenological implications.
Our method can be easily applied to find the full set of
anomaly-free solutions to well defined phenomenological
problems. 

In this work, we look for anomaly-free solutions to SM extensions with an additional $U(1)$ gauge symmetry, giving rise to scotogenic Dirac neutrino masses.
For that purpose, in section~\ref{sec:anomaly} we study the conditions to have a non-anomalous $U(1)$ gauge symmetry.
In particular, we show that the case where the SM is extended with an active Abelian gauge symmetry $U(1)_X$ with non-vanishing generation independent charges and a set of $N'$ singlet chiral fermions, shares the same anomaly-free solutions as an Abelian gauge dark symmetry $U(1)_D$ with $N'+3$ singlet chiral fermions having at least three equal charges.
In section~\ref{sec:Dirac}, we focus on solutions giving rise to scotogenic neutrino masses via the effective Dirac mass operator.
For that purpose, we perform an intensive scan, looking for non-anomalous charge assignments for given number of extra chiral fermions.
We study the cases where the Dirac mass operator, with dimension 5 or 6, is mediated by Dirac or Majorana states, and corresponds to an active $U(1)_X$ or a dark $U(1)_D$ symmetry.
We pay special attention to the solutions where all chiral fermions obtain mass via the spontaneous symmetry breaking (SSB) of the $U(1)$.
Finally, in section~\ref{sec:con} our conclusions are presented.

%%%%%%%%%%%%%%%%%%%%%%%%%%%%%%%%%%%%%%%%%%%%%%%%%%%%%%%%%%%%
%%%%%%%%%%%%%%%%%%%%%%%%%%%%%%%%%%%%%%%%%%%%%%%%%%%%%%%%%%%%
%%%%%%%%%%%%%%%%%%%%%%%%%%%%%%%%%%%%%%%%%%%%%%%%%%%%%%%%%%%%
\section{Anomaly conditions} \label{sec:anomaly}
We consider an extension of the SM with an additional $U(1)_X$ gauge symmetry, and $N'$ right-handed chiral fields $\psi_\rho$ singlets under the SM $SU(3)_c\otimes SU(2)_L\otimes U(1)_Y$ group, with charges $n_{\rho}$ under the $U(1)_X$, where $\rho=1,\cdots,N'$.
Additionally, we assume that the SM right-handed chiral fermions transform under the $U(1)_X$, with charges denoted with the same name of the field.%
\footnote{$Q$ and $L$ are the $X$-charges of the fermion doublets $Q^{\dagger}$ and $L^{\dagger}$, respectively.}
To avoid having an anomalous $U(1)_X$, the three linear anomaly conditions are
\begin{align}
    \left[SU(3)_c\right]^{2}  U(1)_{X} &: \quad [3u+3d] + [3\times 2Q]=0\,,\\
    \left[SU(2)_{L}\right]^{2} U(1)_{X} &: \quad [2L+3\times 2Q]=0\,,\\
    \left[U(1)_{Y}\right]^{2} U(1)_{X} &: \quad \left[(-2)^2e + 3\left(\frac43\right)^2u + 3\left(-\frac23\right)^2d\right] + \left[2(-1)^2L + 3\times 2\left(\frac13\right)^2Q\right] =0\,.
\end{align}
As they only depend on the SM fermions, three of their $X$-charges can be expressed in terms of the other two~\cite{Appelquist:2002mw,Campos:2017dgc,Das:2017flq, Calle:2019mxn}, chosen to be $e$ and $L$, as
\begin{align}\label{eq:sol0}
  u=&-e-\frac23 L\,,& d=& e+\frac43 L\,,& Q=& -\frac13 L\,.
\end{align}
We note that the quadratic anomaly condition in $U(1)_X$ is trivially satisfied. 
However, the mixed gauge-gravitational $\left[ \text{Grav} \right]^{2} U(1)_{X}$ and the cubic $\left[U(1)_{X}\right]^{3}$ anomalies do depend on the extra fermion charges $n_\rho$, and therefore two additional conditions have to be imposed in order to avoid an anomalous $U(1)_X$~\cite{Calle:2019mxn}:
\begin{align}\label{eq:grav}
  \sum_{\rho=1}^{N'}n_{\rho}+3m&=0\,,&  \sum_{\rho=1}^{N'}n_{\rho}^3+3m^3&=0\,,
\end{align}
where $m \equiv e+2L$.
Equation~\eqref{eq:sol0} can be rewritten as
\begin{align}
  u=&\frac{4 L}{3}-m\,,& d=m-\frac{2 L}{3}& \,,& Q=& -\frac{L}{3}\,,& e=& m-2 L\,.
\end{align}
Finally, we note that the SM Higgs must have a $X$-charge
\begin{align}
  h=-e-L=L-m\,,
\end{align}
to guarantee that SM quarks and charged leptons acquire masses through the standard Higgs mechanism.%
\footnote{In particular, if $h=0$ a gauge symmetry with SM-fermion charge $X=m\,(B-L)$ is obtained, where $B-L$ are the baryon-minus-lepton charges.}
Along these lines, we also assume that the singlet chiral fermions $\psi_\rho$ only acquire mass through the SSB of the extra $U(1)_X$ symmetry.
This excludes solutions with vector-like states.
We note that the existence of fields charged under both hypercharge and $U(1)_X$ induce at loop level the kinetic mixing operator $\mathcal{L} \supset \frac{\epsilon}{2} B^{\mu\nu} X_{\mu\nu}$, where $B^{\mu\nu}$ and $X^{\mu\nu}$ are the field strengths related to the $U(1)_Y$ and the extra $U(1)_X$, respectively.
The dimensionless parameter $\epsilon$ depends on the masses of the particles in the loop, as well as their specific charge assignment and the gauge couplings under the two $U(1)$ symmetries~\cite{Holdom:1985ag, Cheung:2009qd, Gherghetta:2019coi}.

It is interesting to note that the conditions in Eq.~\eqref{eq:grav} are completely equivalent to the ones coming from a scenario where the SM is extended with a dark $U(1)_D$ gauge symmetry with $N=N'+3$ right-handed singlet chiral fermions, $N'$ of them with the charges $n_\rho$ and three with charge $m$, and where the SM is invariant (hence a {\it dark} symmetry).
Even if comparable, there is a major technical advantage of the latter approach:
If the SM is extended with and additional dark $U(1)_D$ gauge symmetry (under which it is uncharged), and $N$ right-handed chiral fields singlets under the SM group, the $U(1)_D$ is not anomalous if the Diophantine equations
\begin{align}
 \label{eq:NN3}
 \sum_{\rho=1}^{N}n_{\rho}=0 \qquad \text{and} \qquad \sum_{\rho=1}^{N}n_{\rho}^3=0\,,
\end{align}
coming from the mixed gauge-gravitational $\left[ \text{Grav} \right]^{2} U(1)_D$ and cubic $\left[U(1)_D\right]^{3}$ conditions are fulfilled.
It is well know that the solution of Eq.~\eqref{eq:NN3} is highly nontrivial~\cite{deGouvea:2015pea, Berryman:2016rot, Rathsman:2019wyk, Costa:2019zzy, Batra:2005rh, Costa:2020dph}.
However, for an $U(1)$ symmetry, it can be parametrized as a function of two sets of integers $\ell$ and $k$, with dimensions $(N-3)/2$ and $(N-1)/2$ for $N$ odd, or $N/2-1$ and $N/2-1$ for $N$ even~\cite{Costa:2019zzy}.
We have implemented an official Python package called \texttt{anomalies}\footnote{\url{https://pypi.org/project/anomalies/}} to obtain the solution associated to any set of integers $\ell$ and $k$.

In general, a very large number of solutions can be found for a given $N$.
For example, for $N \le 9$, $\mathcal{O}\left(10^9\right)$ different combinations of $\ell$ and $k$ exist, if one allows charges to reach a maximum absolute value of 30.
However, the number of chiral solutions (i.e., without featuring vector-like states) reduces to around $30\,000$.
We note that nearly half of them have at least a pair of repeated charges, corresponding to a requirement for having at least a couple of RHNs.
This large number of solutions is too large in practice to perform a phenomenological study.
However, a manageable set of solutions can be found by imposing further constraints.
For example, one can look for realizations of the effective Dirac mass operator of dimensions 5 and 6, through scotogenic models mediated by heavy chiral fermions.
This possibility will be explored in the next section.

%%%%%%%%%%%%%%%%%%%%%%%%%%%%%%%%%%%%%%%%%%%%%%%%%%%%%%%%%%%%
%%%%%%%%%%%%%%%%%%%%%%%%%%%%%%%%%%%%%%%%%%%%%%%%%%%%%%%%%%%%
%%%%%%%%%%%%%%%%%%%%%%%%%%%%%%%%%%%%%%%%%%%%%%%%%%%%%%%%%%%%
\section{Dirac scotogenic models} \label{sec:Dirac}
In this section we look for anomaly-free $U(1)_D$ or $U(1)_X$ gauge extensions of the SM, with  $N$ or $N-3$ singlet chiral fermions, respectively, realizing one-loop effective Dirac neutrino mass operators~\cite{Cleaver:1997nj, Gu:2006dc}.
In the two-component spinor notation, they can be written as
\begin{align} \label{eq:nmo56}
    \mathcal{L}_{\text{eff}} = h_{\nu}^{\alpha i} \, \left( \nu_{R\alpha}\right)^{\dagger} \, \epsilon_{ab} \, L_i^a \, H^b \left(\frac{S^*}{\Lambda}\right)^\delta + \text{H.c.},\qquad \text{with $i=1,2,3$}\,,
\end{align}
and $\delta = 1$ or $2$ for dimension 5 (D-5) or 6 (D-6) operators, respectively.
Here $h_{\nu}^{\alpha i}$ correspond to dimensionless induced couplings, $\nu_{R\alpha}$ are at least two RHNs ($\alpha=1,2,\ldots$) with the same $D$- or $X$-charge $\nu$, $L_{i}$ are the lepton doublets with $X$-charge $-L$, $H$ is the SM Higgs doublet with $X$-charge $h=L-m$, $S$ is the complex singlet scalar responsible for the SSB of the anomaly-free gauge symmetry with $D$- or $X$-charge $s=-\nu/\delta$ or $s=-(\nu+m)/\delta$, respectively, and $\Lambda$ is a scale of new physics, which is parametrically the typical mass scale of the new (heavy) states.
In general, after the SSB, a remnant $\mathbb{Z}_{|s|}$ discrete symmetry is left, which can guarantee the stability of a potential DM candidate~\cite{Batell:2010bp,Farzan:2012ev}.
Additionally, we note that the SM Higgs boson can mix with the scalar $S$, after symmetry breaking.
In general, the mixing between these two bosons can lead to a stabilization of the metastable electro-weak vacuum of the SM~\cite{Falkowski:2015iwa}.

%%%%%%%%%%%%%%%%%%%%%%
\begin{figure}
  \def\scl{0.5}
  \def\sepf{4.7cm}
  \centering
  \includegraphics[scale=\scl]{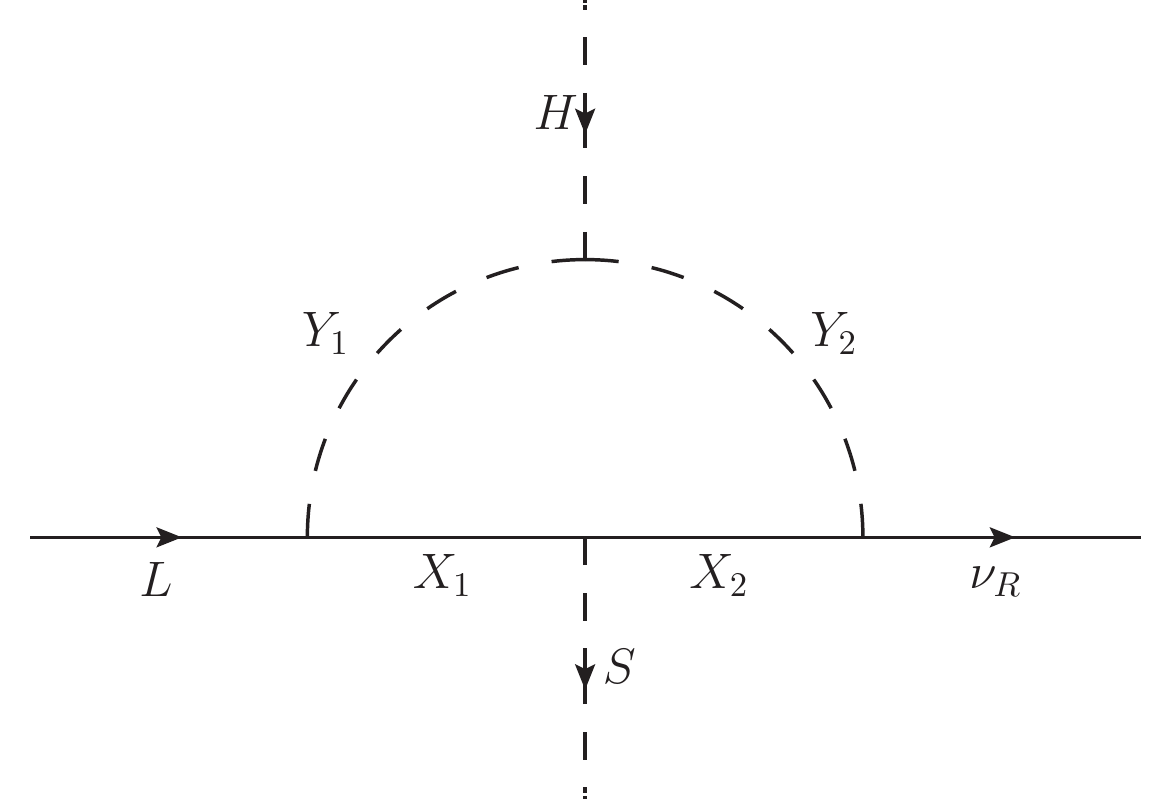}
  \includegraphics[scale=\scl]{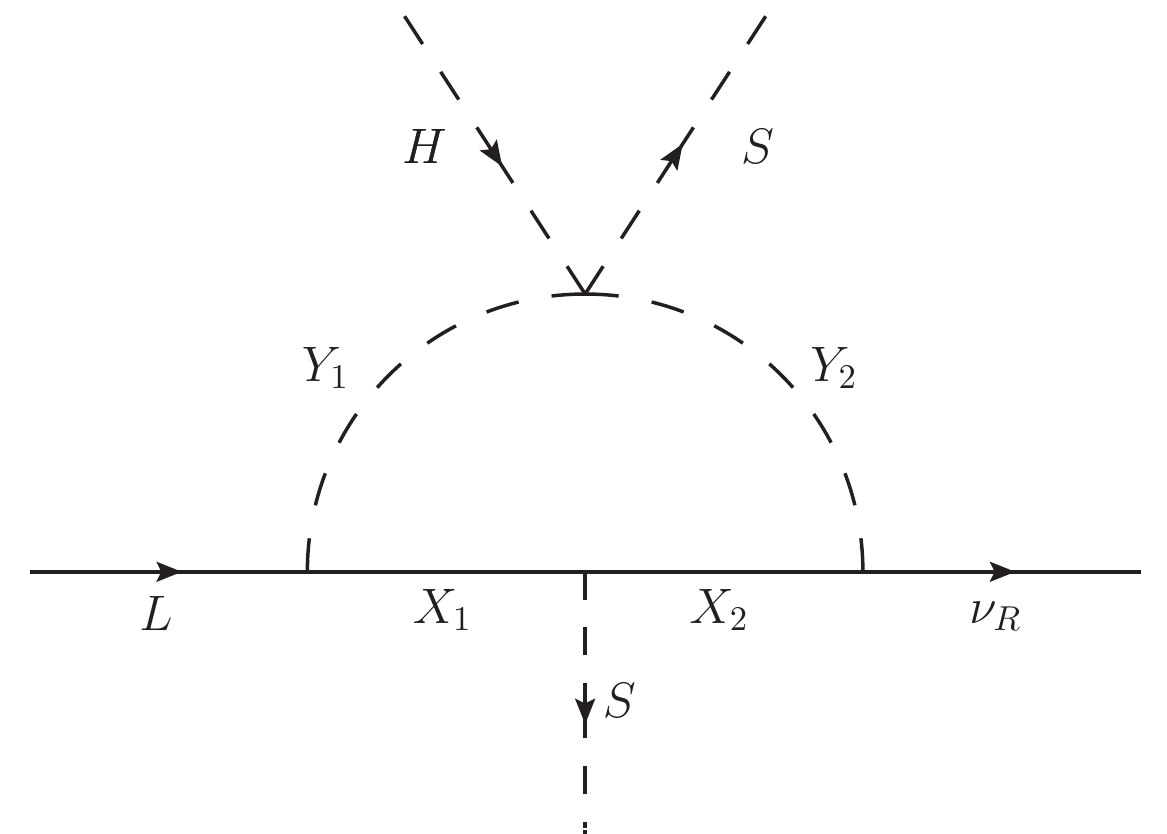}\\
  \quad \texttt{\scriptsize T1-3-E \hspace{\sepf} T1-3-E-D-6} \\
  \vspace{0.5cm}
  \includegraphics[scale=\scl]{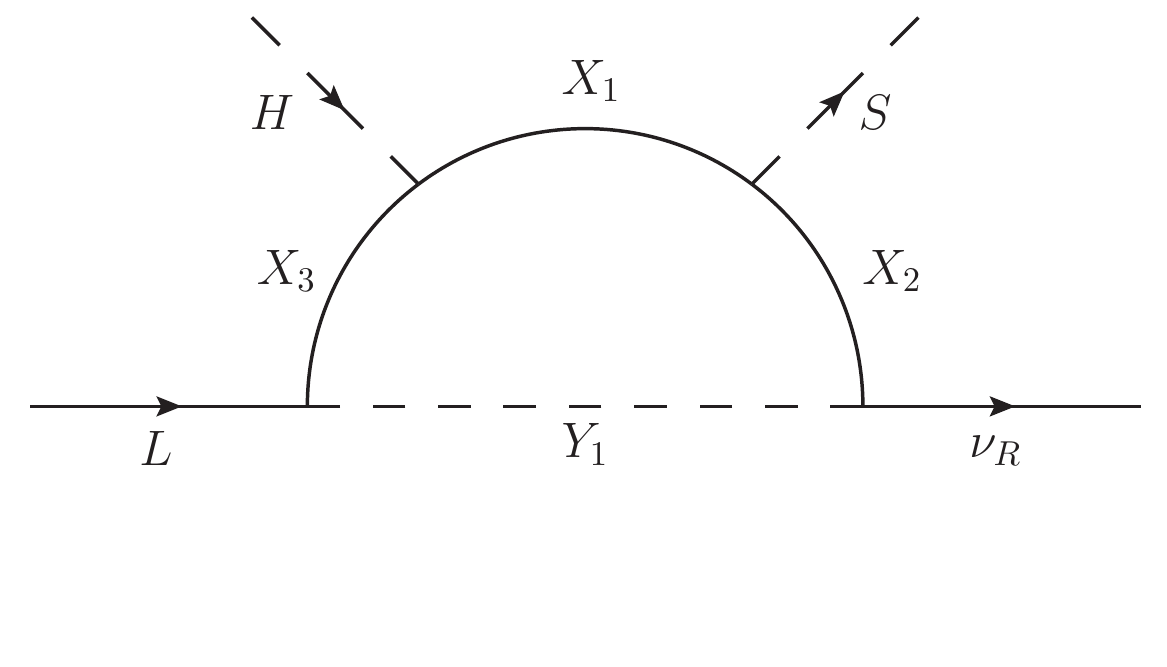}
    \includegraphics[scale=\scl]{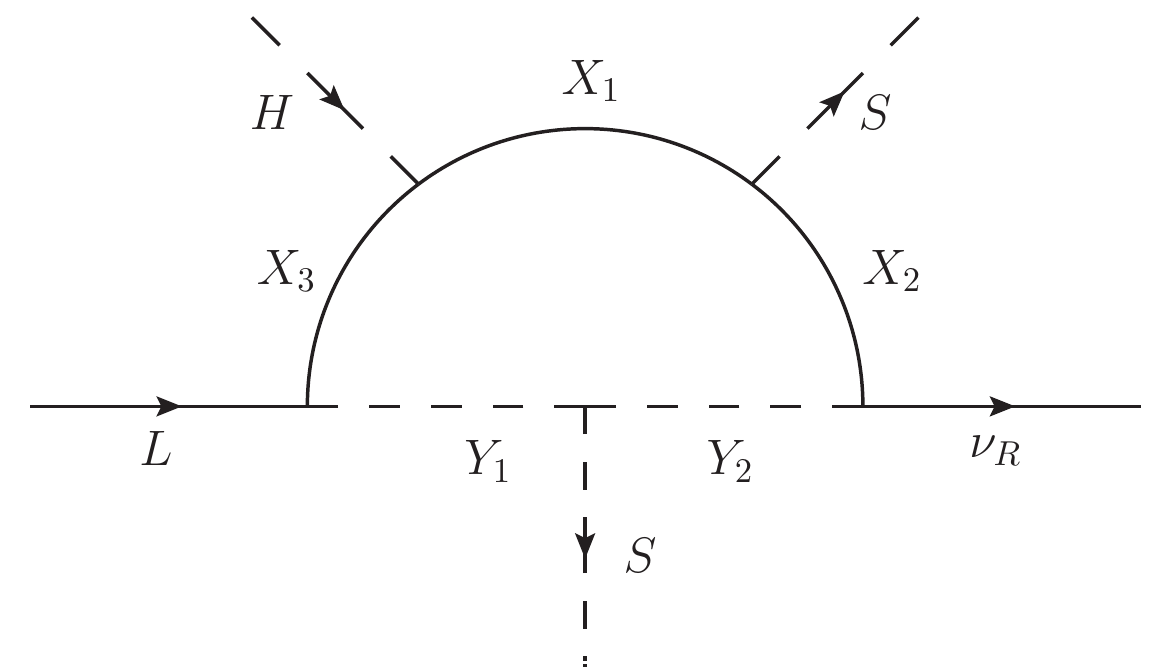}\\
    \quad \texttt{\scriptsize T1-2-A \hspace{\sepf} T1-2-A-D-6}\\
  \vspace{0.5cm}     
  \includegraphics[scale=\scl]{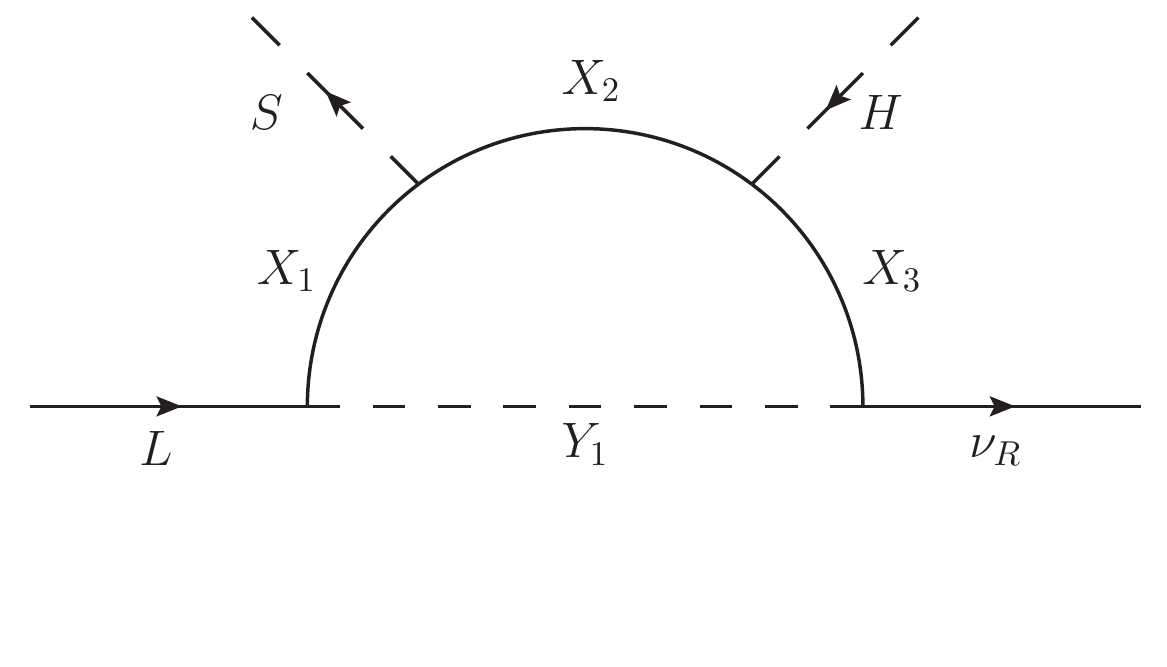}
    \includegraphics[scale=\scl]{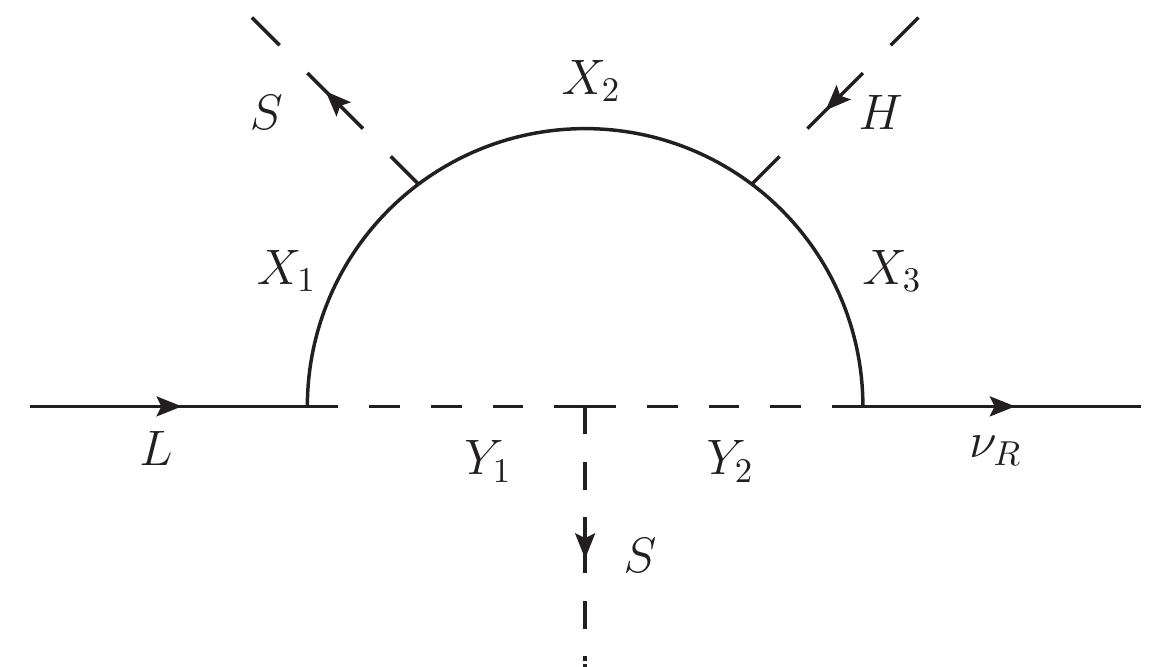}\\
    \quad \texttt{\scriptsize T1-2-B \hspace{\sepf} T1-2-B-D-6}
  \caption{Topologies leading to one-loop Dirac neutrino masses, via the dimension 5 (left panels) or dimension 6 (right panels) operators described in Eq.~\eqref{eq:nmo56}.
  Here we follow the notation of used in Ref.~\cite{Yao:2018ekp}.}
  \label{fig:topo}
\end{figure}
%%%%%%%%%%%%%%%%%%%%%%
The left (right) panels of Fig.~\ref{fig:topo} present the topologies realizing the D-5 (D-6) effective Lagrangian in Eq.~\eqref{eq:nmo56}, if one only allows SSB masses for the singlet chiral fermions.
The different diagrams are labelled following the notation used in Ref.~\cite{Yao:2018ekp}.
Here $X_1$ and $X_2$ correspond to the singlet chiral fermion fields. Without loss of generality, one can choose $X_1$ to be a right-handed field $\chi_R$ with $D$- or $X$-charge $r$. After the SSB of the extra Abelian symmetry, a heavy Majorana or Dirac mediator is generated depending if one chooses $X_2$ to be the same $\chi_R$ or a new $\left(\chi_L\right)^{\dagger}$, with $D$- or $X$-charge $l$.
In the last two rows, $X_3$ is a doublet vector-like fermion with an additional coupling to the Higgs and a singlet chiral fermion.
Finally, $Y_i$ corresponds to either singlet or doublet inert scalars according to the specific vertex. For example, for the topology \texttt{T1-3-E}, $Y_1$ is an inert doublet scalar, while $Y_2$ is an inert singlet scalar.

%%%%%%%%%%%%%%%%%%%%%%%%%%%%%%%%%%%%%%%%%%%%%%%
\begin{figure}
  \def\scl{0.5}
  \def\sepf{4.7cm}
  \centering
  \includegraphics[scale=\scl]{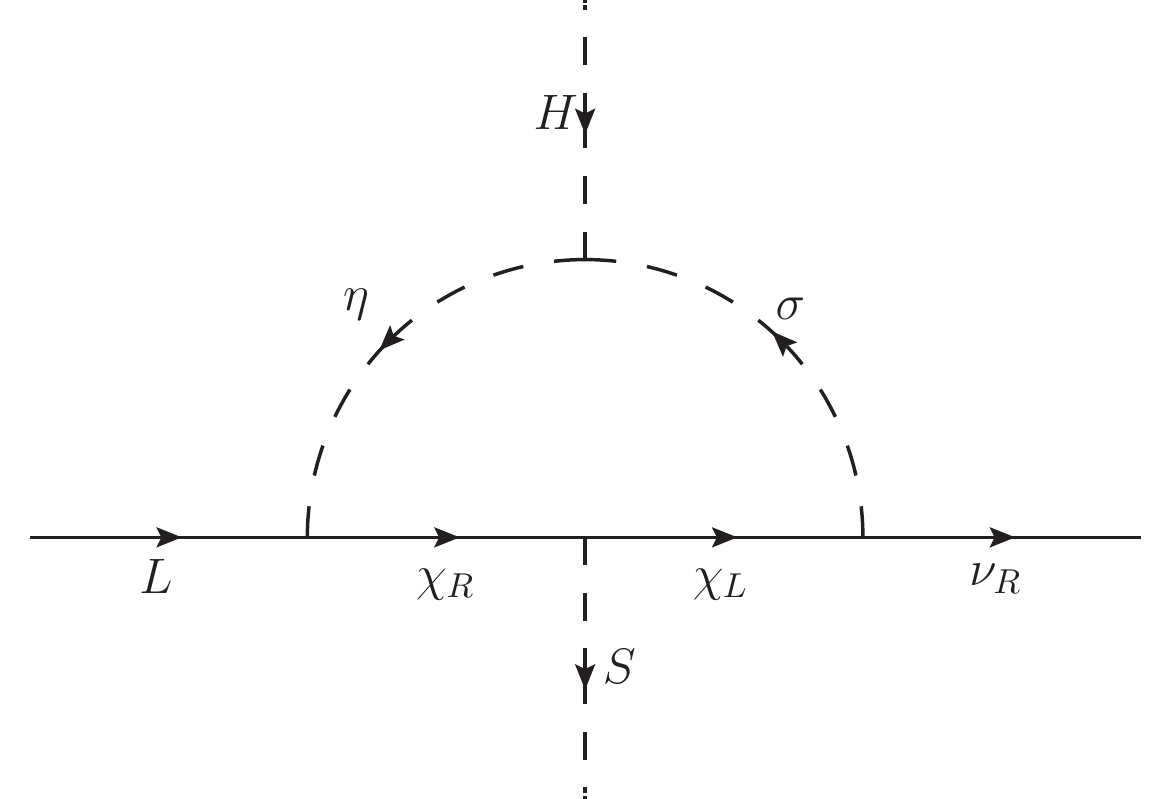}
  \includegraphics[scale=\scl]{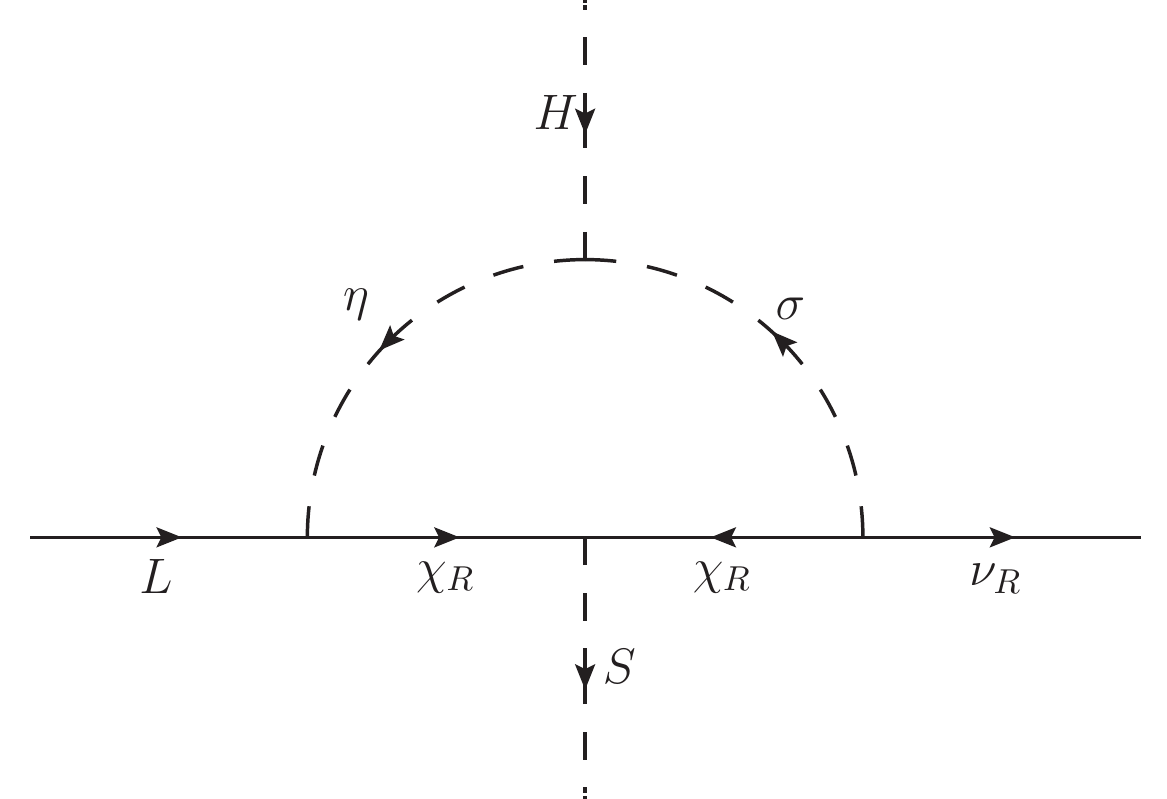}\\
  {\scriptsize \texttt{T1-3-E-D} \hspace{\sepf} \texttt{T1-3-E-M}} \\
  \vspace{0.2cm}
  \includegraphics[scale=\scl]{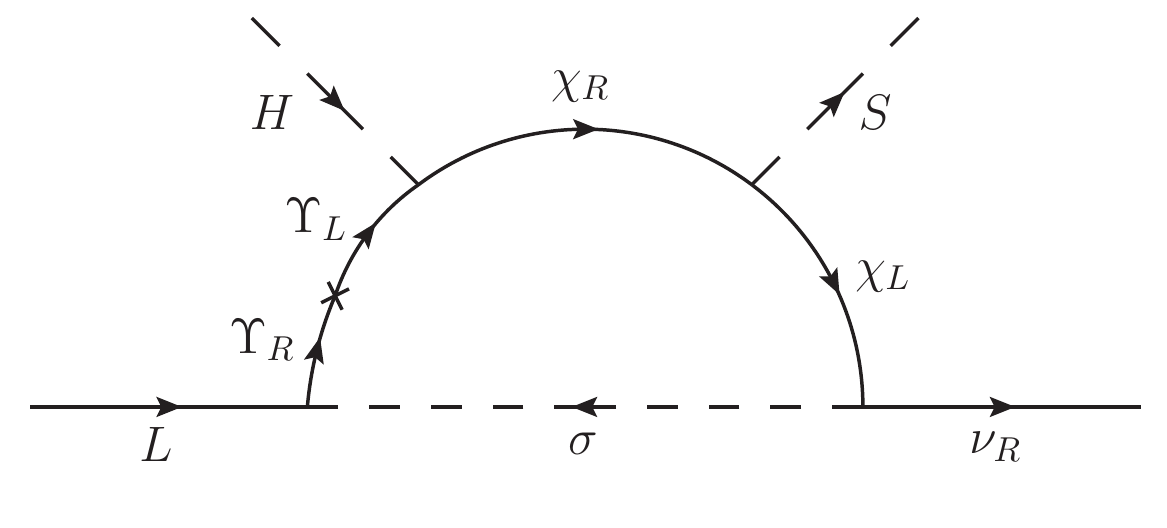}
    \includegraphics[scale=\scl]{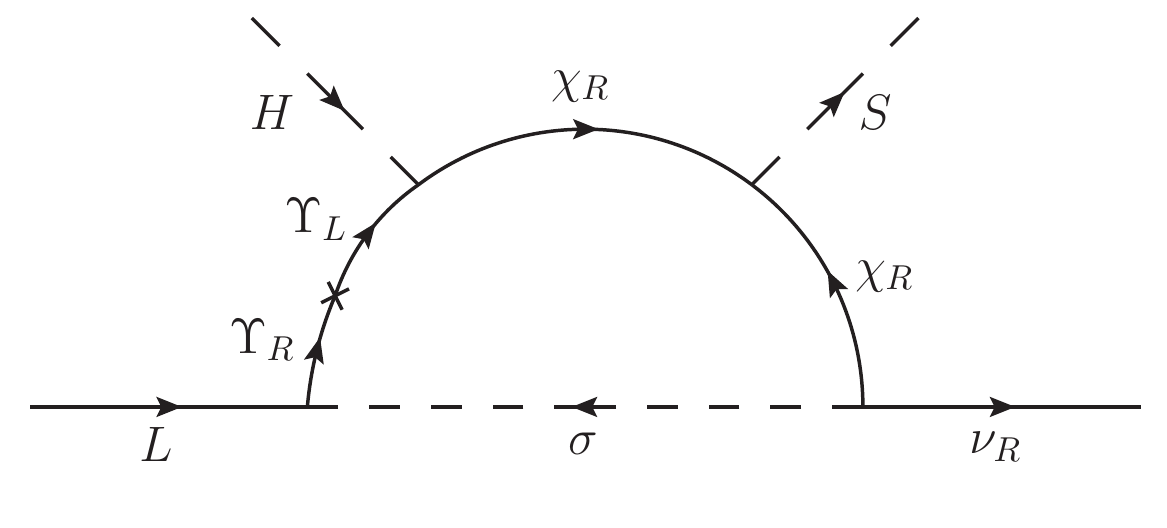}\\
  \vspace{0.2cm} 
  \quad {\scriptsize \texttt{T1-2-A-D} \hspace{\sepf} \texttt{T1-2-A-M}} \\
  \includegraphics[scale=\scl]{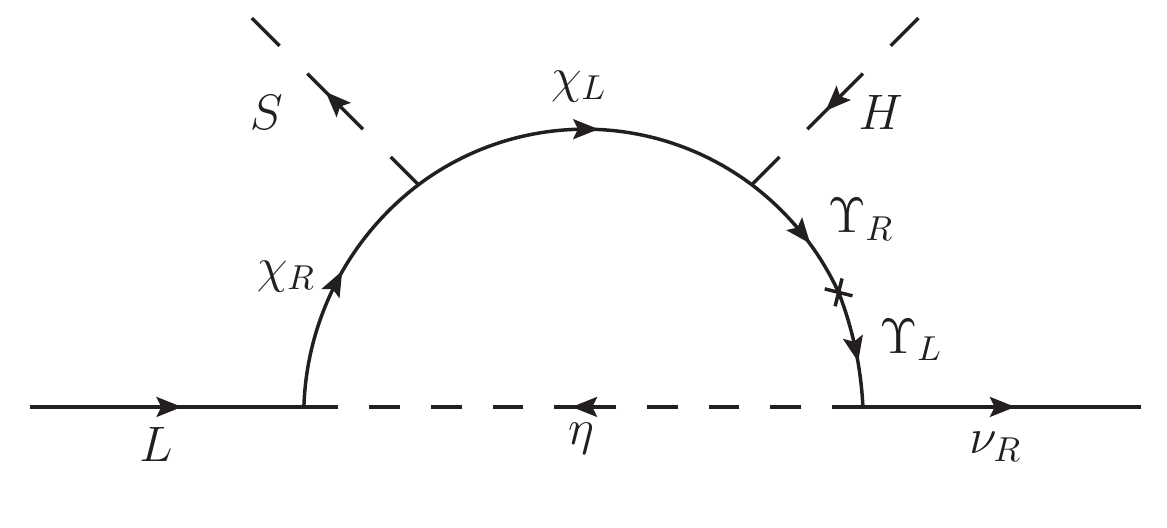}
    \includegraphics[scale=\scl]{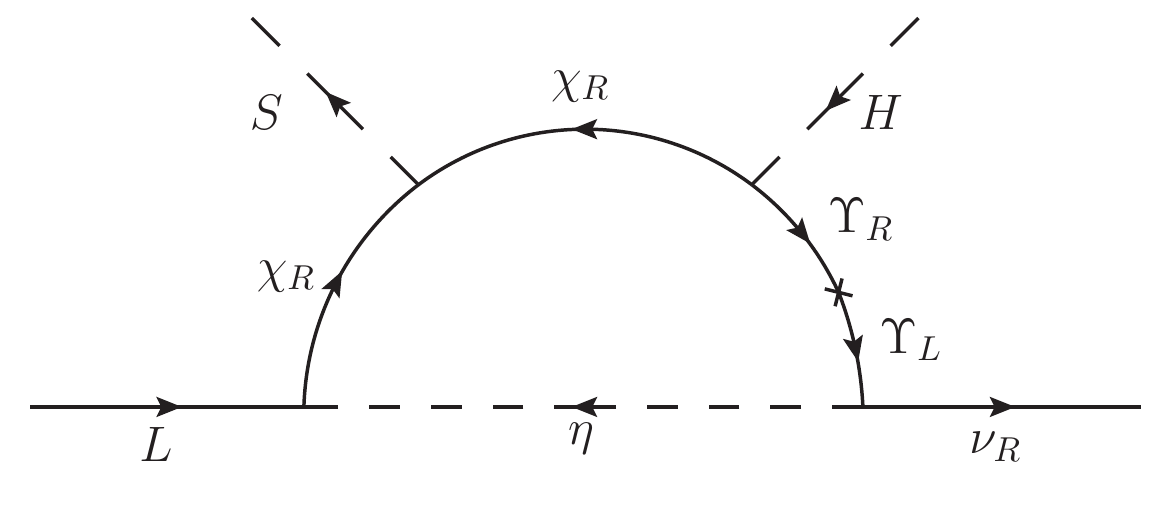}\\
  \vspace{0.2cm} 
  \quad {\scriptsize \texttt{T1-2-B-D} \hspace{\sepf} \texttt{T1-2-B-M}} \\
  \caption{Diagrams for the D-5 realizations of Dirac neutrinos mass with Dirac heavy mediators (left) and Majorana heavy mediator (right). Similar diagrams but with an extra external singlet scalar line, are expected for the D-6 realization}
  \label{fig:toporea}
\end{figure}
%%%%%%%%%%%%%%%%%%%%%%%%%%%%%%%%%%%%%%%%%%%%%%%
The realizations of the D-5 operator featuring heavy Dirac (Majorana) mediators are shown in the left (right) panel of Fig.~\ref{fig:toporea}.
The realizations of the D-6 operators are almost identical, but with an additional extra external singlet scalar line, and are therefore not shown.
For example, the upper left diagram in Fig.~\ref{fig:toporea}, labeled as \texttt{T1-3-E-D}, is realized for a dark $U(1)_D$ symmetry if the flux of the $D$-charges in each vertex satisfy 
\begin{align}
  \eta=&r\,, & s=&r+l\,,& \sigma =&-l-\nu\,, & \sigma=&\eta \,,
\end{align}
and therefore, the fermion chiral $D$-charges obey
\begin{align}
    \nu + l + r = 0\,.
\end{align}
 The scalar $D$-charges can be expressed as a function of $l$  and $r$ as
\begin{align}
    &s = -\nu = r+l\,,  &\sigma = \eta = r\,.
\end{align}
In general, for realizing one-loop Dirac neutrino masses one requires
\begin{equation}
  \label{eq:grlcndtn}
    \nu + \delta\,(l + r) + m = 0\,,
\end{equation}
with $m=0$ for a gauge $U(1)_D$ symmetry, and $l = r$ for realizations with massive Majorana mediators.

As mentioned previously, in order to limit the total number of solutions that cancel the anomaly induced by the additional $U(1)_X$ or $U(1)_D$ symmetries, the following restrictions are taken into account:
\begin{enumerate}
    \item By construction, all new chiral fermions have to be charged under $U(1)_X$ or $U(1)_D$ symmetries, i.e., solutions with vanishing charges are disregarded.
    \item For the chiral fields, the maximal charge allowed (in absolute value) is 30.
    \item Solutions with vector-like fermions are disregarded, i.e., the ones containing two opposite charges.
    \item At least two charges have to be equal.
    Their corresponding fields are identified with the RHNs.
    \item In the case of a $U(1)_X$ symmetry, another set of three equal charges is required.
    \item We restrict ourselves to $N \le 9$ fields, with charges satisfying the two Diophantine conditions in Eq.~\eqref{eq:NN3}, and take the minimal charge (in absolute value) to be positive.
    We note that there are no solutions for $N \leq 5$ with at least two equal charges~\cite{Davoudiasl:2005ks, Nakayama:2011dj}.
    \item The charge assignment may not allow all chiral fields to acquire masses via the SSB.
    We only consider solutions which have at most two massless chiral fields.
    \item We want RHN masses to be generated radiatively.
    That implies that all vertices between $S$, the RHN and the other chiral fields should be forbidden by the symmetries.
    \item A DM candidate should flow in the loop, in order to have a scotogenic solution.
\end{enumerate}

Additionally, we note that for a fixed number of chiral fields and a given topology, different solutions could share the same qualitative behavior.
For example, there are four solutions for a $U(1)_D$ symmetry with Dirac mediators, $N=6$ chiral fields, and neutrino masses generated via the D-5 operator: $(1,  -2,  -3,   5,   5,  -6)$, $(2,  -3, -10,  13,  13, -15)$, $(3,  -4, -21,  25,  25, -28)$ and $(3,  -5, -12,  17,  17, -20)$.
However, they all feature the same phenomenology, i.e., two RHNs and four other massive chiral fermions.
Therefore, in this case only one solution (the one with the smallest charge in absolute value) will be reported in the following.

%%%%%%%%%%%%%%%%%%%%%%%%%%%%%%%%%%%%%%%%%%%%%%%%%%
\begin{table}
  \centering
  \begin{scriptsize}
  \begin{tabular}{LLR|LRR|LLLL}
 \toprule
             &                &            & \multicolumn{3}{c|}{Solutions}   & \multicolumn{4}{c}{$\nu_R$} \\
 &   &   &   &   &   & \multicolumn{2}{c}{$U(1)_D$} & \multicolumn{2}{c}{$U(1)_X$}\\
N&            \ell &              k &  \text{Charges} &  \text{GCD} & \text{Ref.} & \text{Dirac} &  \text{Maj.} & \text{Dirac} & \text{Maj.} \\
\midrule
 \boldsymbol{6} &       (-1, 1) &           (-2, 0) &                   (1, -2, -3, 5, 5, -6) &   1 & \text{\cite{Ma:2021szi}}    &   (\boldsymbol{5}) &                    &                    &      \\ \midrule
              7 &       (-1, 1) &       (-1, 0, -1) &                (1, 2, 2, -3, -3, -3, 4) &   1 & \text{\cite{Ma:2021szi}}    &               (-3) &                    &                    &      \\
              7 &       (1, -1) &      (-2, -5, -4) &                (1, 1, -3, -4, 6, 6, -7) &   1 &     &             (1, 6) &                (6) &                    &      \\
              7 &       (-1, 1) &      (-1, -2, -1) &                (1, 3, -4, 5, -6, -6, 7) &   1 &     &               (-6) &               (-6) &                    &      \\ \midrule
              8 &  (-1, -5, -3) &      (-6, -4, -7) &             (1, 1, 2, 3, -4, -4, -5, 6) &   1 &     &            (-4, 1) &               (-4) &                    &      \\
              8 &   (-1, 2, -2) &        (-7, 4, 0) &             (1, 2, 2, 2, -3, -5, -6, 7) &   1 &     &                (2) &                    &                    &      \\
              8 &     (1, 2, 1) &    (-5, -10, -11) &             (1, 2, 2, 4, -5, -5, -7, 8) &   1 &     &               (-5) &                    &                    &      \\
 {8} &   (1, -3, -2) &  (-4, -9, -5, -3) &             (1, 3, 3, 3, -5, -7, -7, 9) &   1 &     &                    &                    &  ({-7}) &      \\
              8 &    (2, -1, 0) &   (-5, -9, -1, 0) &            (1, 2, 3, 5, -6, -6, -9, 10) &   1 &     &               (-6) &               (-6) &                    &      \\
              8 &   (-1, 0, -1) &       (-2, 1, -1) &           (2, -5, -5, -5, 7, 8, 8, -10) &   1 & \text{\cite{Calle:2018ovc}}  &                (8) &                    &                (8) &      \\
              8 &    (0, -1, 0) &   (-1, -5, -1, 1) &          (1, 1, 1, -5, -7, 11, 11, -13) &   2 &     &                    &                    &               (11) &      \\
              8 &   (-4, -1, 0) &    (-1, 0, -8, 8) &         (3, -4, -5, 8, 8, -11, -12, 13) &   1 &     &                (8) &                (8) &                    &      \\ \midrule
 \boldsymbol{9} &    (-2, 0, 2) &    (-1, 1, 0, -1) &       (1, 1, -4, -5, 9, 9, 9, -10, -10) &   1 & \text{\cite{Wong:2020obo}}   &   (\boldsymbol{9}) &                    &                    &  (1) \\
              9 &   (-4, -5, 3) &   (-2, 0, -1, -2) &        (3, 3, -4, 5, 5, -6, -8, -8, 10) &   1 &     &         (-8, 3, 5) &                    &                    &      \\
              9 &     (5, 0, 1) &    (-1, -2, 0, 2) &    (1, 1, -5, -7, 12, 14, 14, -15, -15) &   3 &     &       (-15, 1, 14) &               (14) &                    &      \\
              9 &    (1, 4, -1) &   (-2, -5, -4, 8) &          (1, 1, 1, 2, 5, -6, -6, -6, 8) &   1 &     &            (-6, 1) &                    &                    &      \\
              9 &    (-1, 0, 1) &   (-1, 1, -2, -1) &       (1, -3, -3, -3, -5, 8, 8, 8, -11) &   1 &     &            (-3, 8) &                    &                    &      \\
              9 &   (-7, -5, 3) &   (-6, -4, -5, 2) &     (1, -2, -2, -4, 7, -9, 11, 11, -13) &   1 &     &           (-2, 11) &               (-2) &                    &      \\
              9 &    (3, -2, 3) &   (-2, -1, -2, 4) &     (4, 4, 4, -5, -9, -10, -10, 11, 11) &   1 &     &                    &                    &          (-10, 11) &      \\
              9 &   (-2, -6, 5) &  (-5, -1, -3, -6) &          (1, 1, 2, 2, 3, -5, -6, -6, 8) &   1 &     &                (2) &               (-6) &                    &      \\
 \boldsymbol{9} &    (-2, 3, 2) &   (-2, -9, -5, 9) &         (1, -2, 3, 4, 6, -7, -7, -7, 9) &   1 &     &  (\boldsymbol{-7}) &                    &                    &      \\
              9 &   (-3, -1, 5) &   (-9, 3, -4, -1) &         (1, 2, -3, 4, -5, -6, 8, 8, -9) &   1 &     &                (8) &                    &                    &      \\
 \boldsymbol{9} &   (-8, -7, 5) &   (-9, 3, -4, -2) &       (1, -2, -2, -2, 5, -7, 8, 9, -10) &   1 &     &  (\boldsymbol{-2}) &  (\boldsymbol{-2}) &                    &      \\
              9 &  (-4, -1, -4) &   (-3, -5, 1, -4) &         (2, -3, 4, 4, 4, -6, -7, -7, 9) &   1 &     &                (4) &                    &               (-7) &      \\
              9 &   (-3, 1, -2) &  (-4, -3, -6, -3) &       (2, -3, -3, -3, -5, 7, 7, 8, -10) &   1 &     &               (-3) &                    &                    &      \\
              9 &    (-3, 6, 5) &   (-1, -6, 2, -7) &      (2, -3, -3, -3, -6, 7, 7, 11, -12) &   2 &     &                    &                    &                (7) &      \\
 \boldsymbol{9} &   (-4, 2, -3) &   (-2, -5, 5, -6) &        (1, 2, -6, -6, -6, 8, 9, 9, -11) &   2 &     &                (9) &                    &   (\boldsymbol{9}) &      \\
              9 &    (1, -1, 2) &   (-2, -1, 0, -2) &      (2, -3, 4, 6, 6, -7, -10, -11, 13) &   1 &     &                (6) &                (6) &                    &      \\
              9 &  (-2, -1, -3) &  (-1, -4, -3, -4) &       (4, 4, 6, 6, -7, -7, -7, -12, 13) &   1 &     &                    &                    &                (6) &      \\
              9 &    (-4, 1, 2) &    (-1, -4, 2, 1) &    (1, -2, -2, -3, -3, -3, 14, 20, -22) &   2 &     &                    &               (-2) &                    &      \\
              9 &     (1, 4, 7) &    (-1, 6, 4, -5) &     (1, -2, -2, 5, -7, -7, 14, 18, -20) &   4 &     &                    &               (-2) &                    &      \\
              9 &    (8, -1, 0) &  (-1, -6, -3, -6) &  (1, 9, -12, -21, -21, 24, 24, 24, -28) &  18 &     &                    &               (24) &              (-21) &      \\
\bottomrule
\end{tabular}
\end{scriptsize}
\caption{Set of charges satisfying the Diophantine equations together with the conditions enumerated in the text, for $N$ extra chiral fermions, featuring Dirac neutrino masses generated by D-5 operators.
The last four columns correspond to the charges for the RHNs in the cases of $U(1)_D$ or $U(1)_X$ symmetries, and Majorana or Dirac mediators.
The solutions without massless chiral fermions are highlighted with a bold font.}
\label{tab:sltns}
\end{table}
%%%%%%%%%%%%%%%%%%%%%%%%%%%%%%%%%%%%%%%%%%%%%%%%%%
\begin{table}
  \centering
  \begin{scriptsize}
  \begin{tabular}{LLR|LRR|LLLL}
 \toprule
             &                &            & \multicolumn{3}{c|}{Solutions}   & \multicolumn{4}{c}{$\nu_R$} \\
 &   &   &   &   &   & \multicolumn{2}{c}{$U(1)_D$} & \multicolumn{2}{c}{$U(1)_X$}\\
N&            \ell &              k &  \text{Charges} &  \text{GCD} & \text{Refs.} & \text{Dirac} &  \text{Maj.} & \text{Dirac} & \text{Maj.} \\
\midrule
              6 &        (-1, -2) &           (-1, 2) &                     (1, 1, 1, -4, -4, 5) &   1 & \text{\cite{Calle:2019mxn, Ma:2021szi}}    &        &   (-4) &                     &                     \\
              6 &         (1, -2) &           (-4, 1) &                   (1, -4, -4, 9, 9, -11) &   3 &     &        &   (-4) &                     &                     \\
              6 &          (2, 1) &       (-2, -1, 0) &                 (1, -4, -8, 14, 14, -17) &   1 &     &   (14) &        &                     &                     \\ \midrule
              7 &      (1, -2, 1) &        (-9, 6, 3) &                 (1, 1, -4, -4, 7, 8, -9) &   1 &     &        &   (-4) &                     &                     \\
              7 &          (3, 1) &       (-1, -5, 7) &                 (2, 2, -4, 7, -8, -8, 9) &   1 &     &    (2) &   (-8) &                     &                     \\
 \boldsymbol{7} &        (-3, -1) &       (-2, -3, 1) &                 (3, 3, 3, -5, -5, -7, 8) &   1 &     &        &        &   (\boldsymbol{-5}) &                     \\
              7 &        (-3, -4) &      (-5, -7, -4) &                (4, 4, 5, -7, -8, -9, 11) &   2 &     &    (4) &        &                     &                     \\ \midrule
              8 &      (1, 2, -2) &        (-7, 3, 0) &        (1, -7, -7, 17, 17, 19, -20, -20) &   6 &     &  (-20) &        &                     &                     \\
              8 &       (1, 2, 1) &    (-5, -10, -11) &              (1, 2, 2, 4, -5, -5, -7, 8) &   1 &     &    (2) &        &                     &                     \\
              8 &      (3, 2, -2) &    (-4, -3, 4, 5) &          (1, 1, -4, -4, -4, 12, 15, -17) &   4 &     &        &   (-4) &                     &                     \\
              8 &      (-1, 2, 4) &    (-4, 2, -3, 0) &         (4, 4, 7, 14, -16, -16, -22, 25) &   4 &     &    (4) &  (-16) &                     &                     \\
              8 &  (-10, -5, -15) &    (-10, -12, 12) &         (5, 5, 5, -17, -27, -27, 28, 28) & 100 &     &        &        &           (-27, 28) &                     \\
              8 &      (3, 1, -3) &    (-12, -14, -4) &         (1, -3, -3, 5, -11, 12, 12, -13) &   1 &     &   (12) &   (12) &                     &                     \\
              8 &     (-2, 0, -1) &      (-4, -3, -2) &           (1, 2, 2, -8, -8, 12, 15, -16) &   2 &     &    (2) &   (-8) &                     &                     \\
              8 &    (-2, -5, -4) &   (-3, -5, -2, 0) &          (2, -3, 7, -8, -8, 11, 14, -15) &   2 &     &   (-8) &   (-8) &                     &                     \\
              8 &      (0, -9, 4) &   (-4, -6, -7, 4) &         (1, -2, -4, -4, -4, 15, 22, -24) &   2 &     &        &   (-4) &                     &                     \\
              8 &     (-1, 0, -1) &       (-9, 1, -1) &          (3, 3, 3, -7, 17, -23, -23, 27) &   4 &     &        &        &               (-23) &                     \\
              8 &       (0, 1, 0) &   (-1, -4, 3, -4) &       (1, -5, -11, 15, -16, 20, 20, -24) &   2 &     &   (20) &   (20) &                     &                     \\ \midrule
 \boldsymbol{9} &      (3, -4, 5) &   (-4, -3, 1, -3) &      (1, -3, 8, 8, 8, -12, -12, -17, 19) &   4 &     &        &        &  (\boldsymbol{-12}) &  (\boldsymbol{-12}) \\
              9 &     (-2, 6, -4) &    (-8, -7, 6, 3) &      (3, 3, 3, 5, -16, 22, -23, -23, 26) &  20 &     &        &        &               (-23) &               (-23) \\
              9 &      (-9, 2, 3) &   (-1, -7, 6, -9) &        (1, -4, 5, 5, -9, -9, -9, 10, 10) &   3 &     &        &        &                     &                 (5) \\
              9 &      (1, 4, -1) &   (-2, -5, -4, 8) &           (1, 1, 1, 2, 5, -6, -6, -6, 8) &   1 &     &   (-6) &        &                     &                     \\
              9 &      (-9, 6, 7) &     (-2, 4, 3, 1) &       (1, 1, 1, 4, -9, -10, -10, 11, 11) &   3 &     &  (-10) &        &           (-10, 11) &                     \\
              9 &    (-3, -2, -4) &   (-1, -9, -7, 4) &       (3, 3, 3, -4, -4, 8, -11, -11, 13) &   2 &     &   (-4) &        &           (-11, -4) &                     \\
              9 &     (-3, 0, -1) &  (-4, -1, -6, -4) &     (2, -3, -3, -8, -9, 12, 12, 14, -17) &   3 &     &   (12) &   (12) &                     &                     \\
              9 &       (4, 6, 4) &   (-3, -4, -3, 5) &   (3, 4, -10, -10, -10, 12, 12, 13, -14) &   2 &     &   (12) &        &                (12) &                     \\
              9 &      (2, 7, -4) &   (-5, -6, 3, -6) &      (1, 1, 1, -4, -4, -11, 18, 26, -28) &  50 &     &        &   (-4) &                     &                     \\
              9 &     (-3, -6, 2) &    (-5, 1, 7, -8) &    (3, -4, -4, -9, -13, 16, 16, 16, -21) &  90 &     &        &   (16) &                (-4) &                     \\
              9 &      (5, -3, 7) &    (-1, 3, 2, -4) &     (5, 7, 7, -8, -15, -15, -15, 17, 17) &  22 &     &        &        &             (7, 17) &                     \\
              9 &     (-6, -3, 5) &   (-4, -2, -6, 8) &     (4, 7, -8, 9, -16, -16, -16, 18, 18) &   4 &     &   (18) &  (-16) &                (18) &                     \\
              9 &      (-2, 1, 2) &    (-6, -8, 7, 6) &        (4, 4, 4, 5, -6, -6, -6, -10, 11) &   2 &     &        &        &             (4, -6) &                     \\
              9 &     (-9, 2, -3) &    (-2, -8, 5, 2) &     (1, -2, -2, -4, -4, -4, 17, 27, -29) &  81 &     &        &   (-4) &                     &                     \\
              9 &     (-5, -4, 0) &   (-2, -1, -4, 4) &    (1, -4, -4, -4, 12, -14, 15, 18, -20) &   2 &     &        &   (-4) &                     &                     \\
              9 &      (2, 4, -2) &    (-1, -4, 9, 3) &      (1, -4, -4, -5, 7, -9, 16, 23, -25) &   2 &     &   (-4) &   (-4) &                     &                     \\
\bottomrule
\end{tabular}
\end{scriptsize}
\caption{Same as Table~\ref{tab:sltns} but for neutrino masses generated via D-6 operators.}
\label{tab:sltnsD6}
\end{table}
%%%%%%%%%%%%%%%%%%%%%%%%%%%%%%%%%%%%%%%%%%%%%%%%%%
The solutions of the Diophantine equations satisfying all the previously enumerated conditions are shown in Tables~\ref{tab:sltns} and~\ref{tab:sltnsD6}, for the D-5 and D-6 operators, respectively.
The solutions for $N$ extra chiral fermions are parametrized as a function of two sets of integers $\ell$ and $k$ (first three columns).
The fourth column shows the charge assignments, whereas the fifth the general common denominator (GCD) of the original solution.%
\footnote{The solutions presented were normalized to have a GCD equal to one.}
The last four columns correspond to the charges for the RHNs in the cases of $U(1)_D$ symmetry with a Dirac mediator, $U(1)_D$ symmetry with a Majorana mediator, $U(1)_X$ symmetry with a Dirac mediator, and $U(1)_X$ symmetry with a Majorana mediator.%
\footnote{A small set of the solutions has already been explored in Refs.~\cite{Babu:2003is, Batra:2005rh, Calle:2018ovc, Calle:2019mxn, Wong:2020obo, Ma:2021szi}.}
We note that even if most of the solutions contain at least one massless chiral fermion, there are few solutions without (highlighted in bold): 6 in the D-5 case and 3 in the D-6 case.
Regarding these solutions without massless fermions, a few comments are in order:
\begin{itemize}
    \item For a $U(1)_D$ symmetry, a new solution for the D-5 operator ($\delta = 1$) with a Dirac mediator and a minimal set of $N = 6$ fermion chiral fields  was found, corresponding to (1, $-2$, $-3$, 5, 5, $-6$).%
    \footnote{This solution was very recently presented in Ref.~\cite{Ma:2021szi}.}
    In this case, there are two RHNs with charges $\nu = 5$.
    The other four chiral fermions combine in pairs ((1, $-6$) and ($-2$, $-3$)) to form a couple of Dirac fermions that obtain mass via the extra scalar $S$ with charge $s = -\nu/\delta = -5$.
    The two Dirac fermions are stable, and give rise to a multicomponent DM scenario.\\
    The solution mediated by a Majorana state is not viable because it will require a fermion with a charge $r = -\nu/2 = -5/2$ which is absent.
    Finally, there cannot be solutions for the $U(1)_X$ symmetry, as there is no charge repeated three times.
    \item The solution with $N = 9$ fermion chiral fields with charges  (1, 1, $-4$, $-5$, 9, 9, 9, $-10$, $-10$) is recovered~\cite{Wong:2020obo}.
    It includes three RHNs with charge 9, together with two pairs (1, $-10$) and one pair ($-4$, $-5$).
    This scenario also features multicomponent DM.
    \item In addition, we found another new solution with $N = 9$: (1, $-2$, 3, 4, 6, $-7$, $-7$, $-7$, 9).
    It includes three RHNs with charge $-7$, and three massive singlet Dirac fermions (1, 6), $(-2,\,9)$, and (3, 4) all of which can be independent DM particles.
    \item The solution with $N = 9$ corresponding to (1, $-2$, $-2$, $-2$, 5, $-7$, 8, 9, $-10$), contains three RHNs with charge $-2$. The condition in Eq.~\eqref{eq:grlcndtn} is fulfilled for the Dirac mediator $(-7,\,9)$ and for the Majorana mediator associated with the chiral fermion of charge $1$. Both of them acquire masses from the SSB of the scalar singlet with charge $2$. This scalar also generates a mixed sector of three chiral fermions $(5,\,-7,\,9)$, and a massive Dirac fermion $(8,\,-10)$.  
    \item The last solution for $N=9$ corresponds to an $U(1)_X$ symmetry and a Dirac mediator, and has the charge assignment: (1, 2, $-6$, $-6$, $-6$, 8, 9, 9, $-11$).
    It includes two RHNs with charge $9$, and four chiral fermions that pair like (1, 2) and (8,  $-11$) to get a mass via the scalar $S$ with charge $-3$.
    \item Let us consider now the D-6 operators.
    For a $U(1)_X$ symmetry with Dirac mediators there is a new solution with $N = 7$ corresponding to (3, 3, 3, $-5$, $-5$, $-7$, 8).
    It includes two RHNs with charge $-5$, and an extra scalar with charge 1 that gives mass to an extra singlet Dirac fermion ($-7$, 8) which is also a DM candidate.
    \item There is another solution with $N = 9$ given by (1, $-3$, 8, 8, 8, $-12$, $-12$, $-17$, 19).
    It includes two RHNs with charge $-12$, an extra scalar with charge 2 which gives mass to the Majorana pair (1, $-3$) and the Dirac pair ($-17$, 19) as two independent DM candidates.
\end{itemize}

All other solutions presented in Tables~\ref{tab:sltns} and~\ref{tab:sltnsD6} have either one or two massless chiral fermions.
They can be either extra relativistic degrees of freedom, or additional DM candidates if they acquire mass from another mechanism.

\begin{figure}
        \centering
        \includegraphics[scale=0.5]{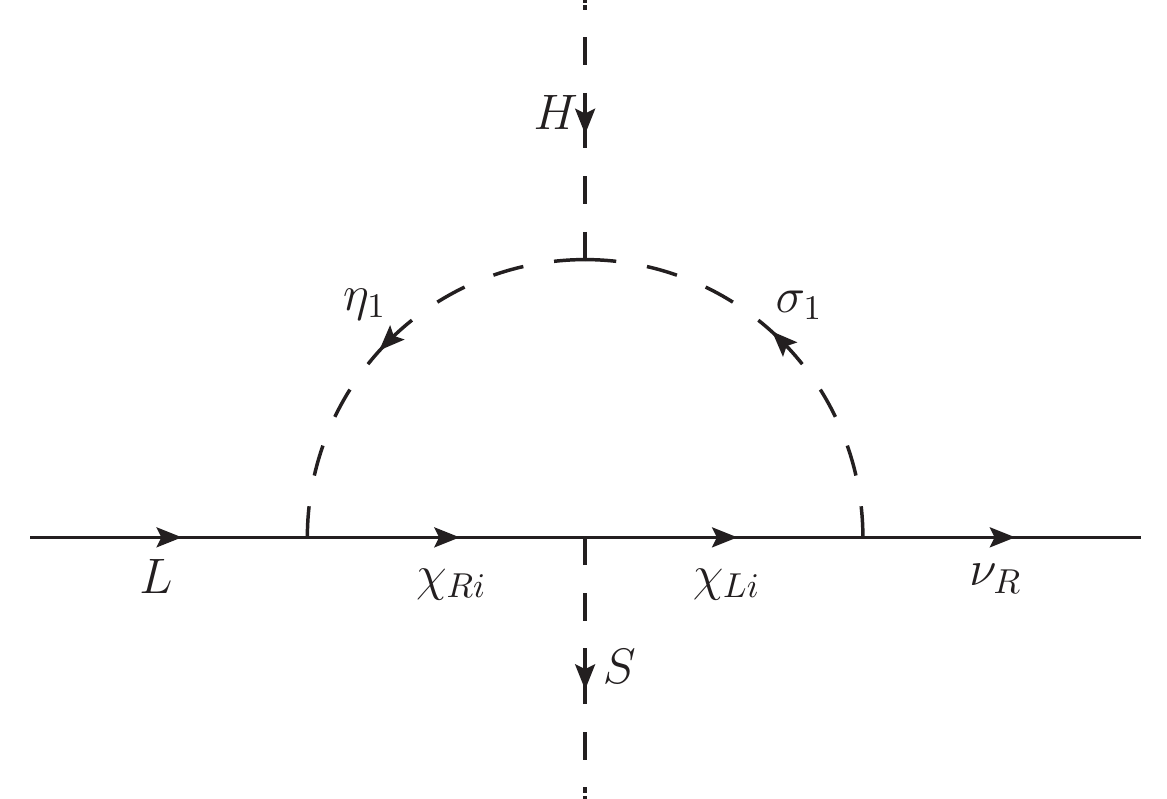}
        \includegraphics[scale=0.5]{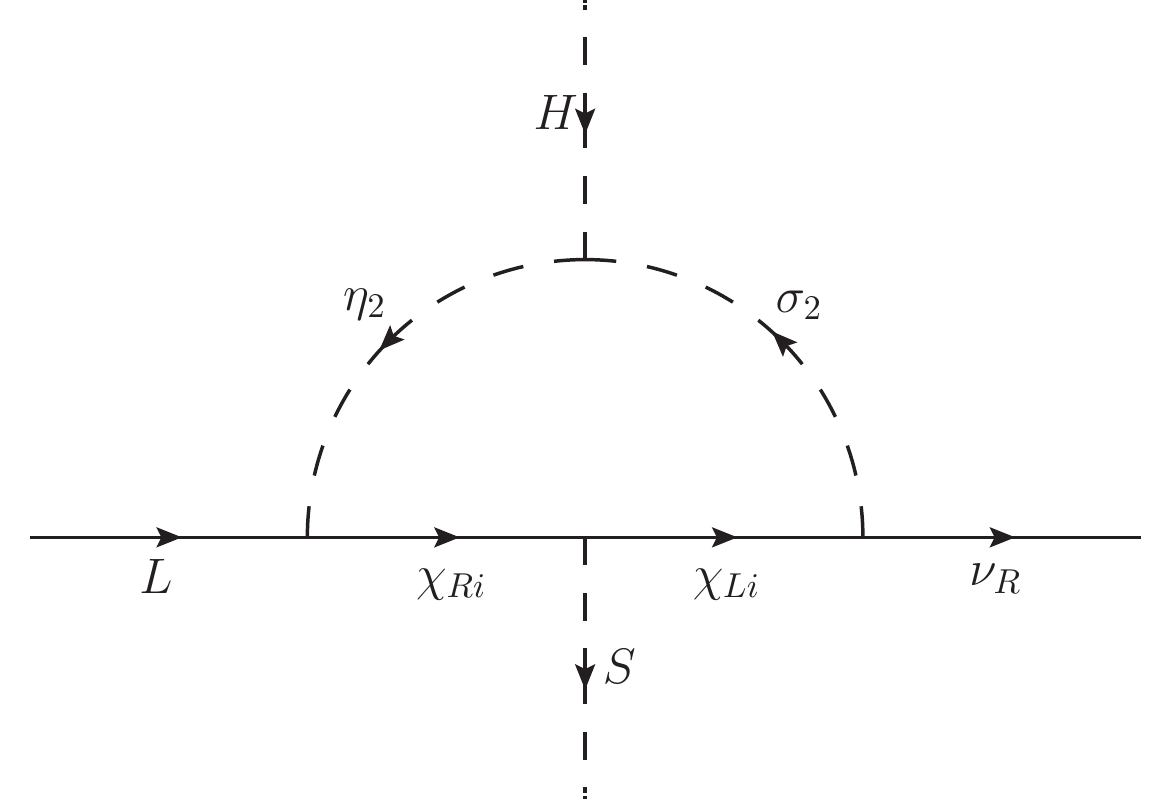}\\
        {\scriptsize \quad \texttt{T1-3-E-D-I} ($i=1$) or \texttt{T1-3-E-D-II} ($i=2$) \hspace{0.5cm} T1-3-E-D-I ($i=1$) or T1-3-E-D-II ($i=2$) }
        \caption{Possible contributions to the Dirac neutrino mass matrix.}
        \label{fig:T1-3-E-I}
\end{figure}

Finally, the one-loop topologies must be realizations of the effective operators of D-5 or D-6 in Eq.~\eqref{eq:nmo56}, with a sufficiently rich $h_{\nu}^{\alpha i}$ structure to explain the full neutrino oscillation data.
That can be guaranteed by having a rank 2 Dirac neutrino mass matrix, via the inclusion of a proper set of inert scalars for each solution.
For example, for the first solution in Table~\ref{tab:sltns} corresponding to the charges (1, $-2$, $-3$, 5, 5, $-6$), three possibilities for the charges of the scalar fields participating in the Dirac neutrino mass loop can be chosen, as shown in the in Table~\ref{tab:sltn1}.
%%%%%%%%%%%%%%%%%%%%%%%%%%%%%%%%%%%%%
\begin{table}
  \centering
  \begin{tabular}{l|c|cccc|cccc|c}
    \toprule
    Field          & $\nu_{R\alpha}$ & $\chi_{R1}$  & $\left( \chi_{L1} \right)^{\dagger}$ & $\eta_1$ & $\sigma_1$ & $\chi_{R2}$  & $\left(\chi_{L2}\right)^{\dagger}$ & $\eta_2$ & $\sigma_2$ & $S$\\
    \midrule
    \texttt{T1-3-E-D-(I-II)}& $5$     & $-2$        & $-3$        & $-2$    &  $-2$      &  $1$         & $-6$       &    $1$      & $1$     & $-5$\\
    \texttt{T1-3-E-D-(I-I)}& $5$     & $-2$        & $-3$        & $-2$    &  $-2$      &  $1$         & $-6$       &    $-2$      & $-2$     & $-5$\\
    \texttt{T1-3-E-D-(II-II)}& $5$     & $-2$        & $-3$        & $1$    &  $1$      &  $1$         & $-6$       &    $1$      & $1$     & $-5$\\
   \bottomrule
  \end{tabular}
  \caption{Possible charge assignments to obtain a light Dirac neutrino mass matrix of rank 2, for the first solution of Table~\ref{tab:sltns}, i.e., (1, $-2$, $-3$, 5, 5, $-6$). In the row
   \texttt{T1-3-E-D-(I-I)} [\texttt{T1-3-E-D-(II-II)}] 
    the second [first] heavy Dirac fermion, from $\left( \chi_{L2} \right)^{\dagger}\chi_{R2}S^{*} $ [$\left( \chi_{L1} \right)^{\dagger}\chi_{R1}S^{*} $], does not participate directly in the neutrino-loop.}
  \label{tab:sltn1}
\end{table}
%%%%%%%%%%%%%%%%%%%%%%%%%%%%%%%%%%%%%
The labels in parentheses in the first column refer to the diagrams in Fig.~\ref{fig:T1-3-E-I}.

Before concluding, we note that the presence of $\eta_i$,%
\footnote{Also the vector-like fermion doublet in the realizations of topologies without $\eta$ (\texttt{T1-2-A-D} and \texttt{T1-2-A-M}) in Fig.~\ref{fig:toporea}.}
through the first vertex from left to right in the diagrams of Fig.~\ref{fig:T1-3-E-I}, i.e.
\begin{align}
    \mathcal{L}\supset y_{1j}\left(\chi_{R1}\right)^\dagger\epsilon_{ab}L^a_j \eta^b_1+\text{H.c.}\,,
\end{align}
imply the presence of lepton flavor violating processes, as for example $l_j\to l_k\gamma $, which  
 is induced at one-loop level and mediated by the charged scalar $\eta_1^+$. 
 By using the current experimental constraint on $\operatorname{Br}(\mu\to e\gamma)<5.7\times 10^{-13}$ at $90\%$ confidence level~\cite{Adam:2013mnn}, upper bounds can be established for the product of Yukawas couplings like
 $|y_{12}y_{11}^*|$~\cite{Calle:2019mxn}.

%%%%%%%%%%%%%%%%%%%%%%%%%%%%%%%%%%%%%%%%%%%%%%%%%%%%%%%%%
%%%%%%%%%%%%%%%%%%%%%%%%%%%%%%%%%%%%%%%%%%%%%%%%%%%%%%%%%
%%%%%%%%%%%%%%%%%%%%%%%%%%%%%%%%%%%%%%%%%%%%%%%%%%%%%%%%%
\section{Conclusions} \label{sec:con}
Even if neutrino experimental data is compatible with both Majorana and Dirac neutrino masses, most of the proposals in the literature assume that neutrinos are Majorana.
Having a Dirac neutrino requires the addition of singlet RHNs trivially charged under the SM gauge group, and an extra symmetry spontaneously broken by a new Higgs field.
Additionally, if the symmetry forbids the tree-level contribution to neutrino masses, a Dirac-seesaw mechanism can be implemented at loop level, avoiding therefore Yukawa coupling to be too small.
Such symmetry can be either a dark $U(1)_D$ under which the SM fields are all singlets, or an active $U(1)_X$ if the SM transforms under its action.
Finally, a non-anomalous theory requires the introduction of a set of singlet fermions, with well defined charges under the new symmetry.

Studies on one-loop Dirac neutrino masses have typically focused on finding specific anomaly-free solutions for this two kinds of symmetries.
In the present work, a complete set of relevant anomaly-free solutions to the general problem of the generation of Dirac neutrino masses at one-loop with chiral singlet fermions has been presented.
In particular, we restricted the analysis to solutions satisfying a set of general conditions enumerated in the text.
In particular, we focused on scotogenic solutions, i.e., the ones with a DM candidate flowing in the loop.
Each of the presented solutions leads to a unique model with specific phenomenological implications.
The full set of solutions is shown in Tables~\ref{tab:sltns} and~\ref{tab:sltnsD6}.

We found 32 and 34 sets of charges that realize the one-loop effective Dirac neutrino mass operator for dimensions 5 and 6, respectively.
Solutions corresponding to both $U(1)_D$ and $U(1)_X$ gauge symmetries, and Dirac or Majorana mediators were analysed.
We emphasize that even if most of the solutions contain one or two massless chiral fermions, there are few solutions (6 and 3 for dimensions 5 and 6, respectively) with all extra fermions getting mass via the spontaneous symmetry breaking of the new Higgs field.
The massless fermions can either contribute to the relativistic degrees of freedom $\Delta N_\text{eff}$ in the early universe~\cite{Calle:2019mxn}, or acquire masses after the introduction of an extra singlet scalar, becoming independent DM candidates~\cite{Bernal:2018aon}.

Finally, we note that the methodology presented can be easily applied to find the full set of anomaly-free solutions to a well defined phenomenological problems.
Additionally, some particular models presented here are the subject of a detailed phenomenological study in an ongoing study~\cite{futuro}.

%%%%%%%%%%%%%%%%%%%%%%%%%%%%%%%%%%%%%%%%%%%%%%%%%%%%%%%%%
%%%%%%%%%%%%%%%%%%%%%%%%%%%%%%%%%%%%%%%%%%%%%%%%%%%%%%%%%
\section*{Acknowledgments}
The authors thank Óscar Zapata for very useful discussions.
NB received funding from Universidad Antonio Nariño grants 2019101 and 2019248, the Spanish FEDER/MCIU-AEI under grant FPA2017-84543-P, and the Patrimonio Autónomo - Fondo Nacional de Financiamiento para la Ciencia, la Tecnología y la Innovación Francisco José de Caldas (MinCiencias - Colombia) grant 80740-465-2020.
The  work  of  DR  is  supported  by  Sostenibilidad  UdeA,  the  UdeA/CODI  Grant2017-16286,  and  by  COLCIENCIAS  through  the  Grant  1115-7765-7253.
This project has received funding/support from the European Union's Horizon 2020 research and innovation programme under the Marie Skłodowska-Curie grant agreement No 860881-HIDDeN.

\bibliographystyle{apsrev4-1long}
\bibliography{biblio}

\end{document}